\documentclass[structabstract]{aa} 
\usepackage{graphicx}
\usepackage{txfonts}

\def\bp{$\beta$~Pic\ }
\def\bpic{$\beta$~Pic\ }

\def\L'{m$_{L'}$\ }

\def\mic{$\mu$m}
\def\pm{$^{+}_{-}$}
\def\deg{$^{\circ}$}

\begin{document}

\title{{ \bp b position relative to the Debris Disk }\thanks{Based on observations collected at the
European Southern Observatory, Chile, ESO; run 086.C-0341(A).}}

\author{
A.-M.~Lagrange \inst{1}
 \and
A. Boccaletti \inst{2}
\and
J. Milli \inst{1}
\and
G. Chauvin \inst{1,3}
\and
M Bonnefoy \inst{3}
\and
D. Mouillet \inst{1}
\and
J.C. Augereau \inst{1}
\and
J. H. Girard  \inst{5}
\and
S. Lacour \inst{2}
\and
D. Apai \inst{4}
}

\offprints{A.-M. Lagrange}

\institute{
  Institut de Plan\'etologie et d'Astrophysique de Grenoble,
  Universit\'e Joseph Fourier, CNRS, BP 53, 38041 Grenoble, France
  \email{anne-marie.lagrange@obs.ujf-grenoble.fr}
\and
LESIA-Observatoire de Paris, CNRS, UPMC Univ. Paris 06, Univ. Paris-Diderot, 92195, Meudon, France 
\and
Max Planck Institut fur Astronomie K\"onigstuhl 17, D-69117 Heidelberg, Germany
\and
Department of Astronomy and Department of Planetary Sciences, The University of Arizona, 933 N Cherry Avenue, Tucson, AZ 85718, USA
\and
ESO Vitacura
Alonso de Cordova 3107, 
Vitacura, Casilla 19001, 
Santiago de Chile 19, 
Chile
}

\date{Received date / Accepted date: 16 01 2012}

\abstract
{  We detected in 2009 a giant, close-by  planet orbiting \bp, a young star surrounded with a disk, extensively studied for more than 20 years. We showed that if located on an inclined orbit, the planet could explain several
 peculiarities of $\beta$ Pictoris\ system. However, the available data did not permit to measure the inclination of \bp b with respect to the disk, and in particular to establish in which component of the disk - the main, extended disk or the inner inclined component/disk-, the planet was located.  Comparison between the observed planet position and the disk orientation measured on previous imaging data was not an option because of potential biases in the measurements.}
{Our aim is to measure precisely the planet location with respect to the dust disk using a single high resolution image, and correcting for systematics or errors that degrades the precision of the disk and planet relative position measurements.}
{We gathered new NaCo data at Ks band, with a set-up optimized to derive simultaneously the orientation(s) of the disk(s) and that of the planet. }
{We show  that the projected position of \bp b is above the midplane of the main disk. With the current data and knowledge on the system, this implies that \bp b cannot be located in the main disk. The data rather suggest the planet being located in the inclined component. }
{}


\maketitle

\section{Introduction}
Understanding planetary systems formation and evolution has become one
of the most exciting challenges in astronomy, since the imaging of a resolved debris disk around the young star $\beta$ Pictoris, in the 80's and the discovery of the
first exoplanet around the solar-type star 51~Peg in the
90's. While more than 500 planets (mostly giants, hereafter GPs) closer than a few AU have been identified with
radial velocity (RV) and transit techniques, very few have been imaged
and definitely confirmed around stars, at separations comparable to those of our solar system giants (\cite{marois08}, \cite{lagrange10}). The  planets imaged so far orbit young stars; indeed the young planets are still hot and the planet-star contrasts are compatible with the detection limits currently achievable, in contrast with similar planets in older systems (we exclude here planetary mass objects detected around brown dwarfes, whose origins are still debated). Noticeably, the stars are of early-types, and surrounded by debris disks, i.e. disks populated at least by small grains with lifetimes so short that they must be permanently produced, probably by destruction (evaporation, collisions) of larger solid bodies. 

Apart from these still rare cases of imaged planets, several such debris disks have now been resolved at optical or near-IR wavelengths, with sometimes peculiar structures (rings, gaps) that could indicate the presence of yet unseen planets. These debris disks, and especially those with already imaged planets, are ideal places to study planet-disk interaction and early ages. Among these systems, the young ($12^{+8}_{-4}$~Myr; \cite{zuckerman01}) and close ($19.3 ^{+}_{-}0.2$~pc; \cite{crifo97}) disk around the A5V star $\beta$ Pictoris\, has been considered as a prototype of young planetary systems. Recently, we detected with NaCo 
(\cite{lenzen03}; \cite{rousset03}) on the Very Large Telescope a companion to \bp orbiting between 8 and 15 AU from the star (\cite{lagrange09}). Using Lyon's group models (\cite{baraffe03}) and the observed L' magnitude, we derived from the available photometry a temperature 
of $\sim 1500$~K and a mass of 9\pm 3 ~MJup. These parameters were later confirmed by new observing techniques/wavelengths at 4.0 \mic\  (resp. \cite{quanz10}), and  Ks band (\cite{bonnefoy11}). \footnote{ Recently, \cite{currie11} published a detection at M band. The photometric information is however to be taken with extreme care as, due to the lack of unsaturated \bp PSF image (necessary to estimate the flux level of the planet in the saturated images), the photometric reference used had been taken 2 years prior to the actual observations, which we believe is not well adapted in the case of ground based imaging, moreover at IR.}
 As the brightness-mass relationships predicted by these models are still influenced at young ages by uncertain initial conditions, we furthermore used RV data to directly constrain its true mass to be less than $\simeq$ 10-25 MJup  (\cite{lagrange11}),  for orbital separations of 8-12 AU. Additional astrometric data obtained in 2010-2011 show finally that the planet semi-major axis is in the range 8-10 AU (\cite{chauvin11}), and its eccentricity less than 0.12. $\beta$ Pic b is today the closest planet ever imaged around a star.
 
Two particularities of the \bp disk are 1) the warping of its inner (less than 80 AU) part, observed with HST and from the ground with adaptative optics since the late nineties (\cite{mouillet97}; \cite{heap00}; \cite{goli06}, \cite{bocca09}), and 2) several asymmetries  in more external parts of the disk observed earlier-on (\cite{kalas95}). We attributed the observed warp to the gravitational perturbation of a massive body located on an inclined orbit, on a disk of planetesimals and the outer asymmetries to the  distribution of the small dust released by collisions among the perturbed planetesimals, and immediately blown away by the star radiation pressure (\cite{mouillet97}; \cite{jca01}). \cite{goli06} interpreted this inner part as a secondary, inclined disk\footnote{It is beyond the scope of the present paper to discuss the origin of the inner warp. In the following, we will refer to this inner part as the warped component or inclined disk}. We showed in \cite{lagrange10} that  given its observed properties and the earlier modeling results, if \bp b was indeed located in the warped part of the disk (or close to it), several of these asymmetries could be accounted for.  
However, the error bars on \bp b  position with respect to the star, and in particular, its position angle (hereafter PA) did not allow to determine whether it is located within the main, outer disk or the inner, warp component. Indeed, both components are separated by only 2-5\deg, very close to the uncertainty of the measured projected PA of the planet. This rather large uncertainty on the planet PA is in part due to the fact that the star center positions in the heavily saturated images that we use to reach the high contrast necessary to detect planets cannot be precisely measured. Indeed, the star center position is obtained from a fit on the low-flux level  wings of the saturated image, which happens to be very variable in shape and present important degrees of asymmetries that, we believe, cannot be corrected for (see below).  
 Recently, \cite{currie11} claimed that \bp b was located in the main disk rather than in the inclined/warped disk.  However, we believe these results have to be taken with peculiar  care. Indeed, the data were calibrated using  platescales and orientations measured on data taken more than one month (L' data) and almost one year (M'  data) after the data presented in their paper. Second, the uncertainties associated to these measurements, between 1.2 and 1.9 degrees, are only the uncertainties associated to the planet centroiding estimates. We show in the following that this uncertainty is by far not the largest source of uncertainty. Third, the comparison between the planet and disk PA relies on a PA of the Main disk measured in low spatial resolution data (\cite{kalas95}) taken years before, and does not take into account any error bars on the disk PA, while revised values were published by our group recently (\cite{bocca09}). Fourth, the authors did not discuss the fact that the observed position angle of the planet with respect to the disk is affected by possible projection effects if, as proposed earlier, the disk(s) is(are) inclined with respect to the line of sight. 
Following Currie et al results, \cite{dawson11} developed dynamical simulations involving 3 scenarios: 1)  \bpic b is in fact on an inclined orbit as described above, responsible for the warp, 2) \bpic b is orbiting within the main disk, and an additional planet is responsible for the warp, and 3) \bpic b was initially on an inclined orbit, responsible for the warp, and then moved back into the main disk through inclination damping. Interestlingly, the second scenario was found to be not viable. 

Accurate measurements of the disk or planet absolute PA are not straightforward and require precise calibrations of detector orientation. Aware of these uncertainties, we decided to obtain new data of the \bp system that would allow to measure {\it directly} the position of the planet with respect to the disk, using data that would show at the same time the main and warped component, as well as \bp b, so as to mimimize as much as possible instrumental systematics and associated uncertainties. In Section 2, we describe the data, the reduction procedures, and we provide the resulting images. We present in Section 3 (resp. Section 4) the methods used to measure the disk (resp. planet) orientations and the associated results.  It has to be noticed that as we are interested in finding the relative position of \bp b with respect to the disk, we do not a priori need to measure the planet and disk {\it absolute} PAs and associated uncertainties (detailed in Appendix A and B). In particular, the planet position relative to the disk will not be influenced by systematics that affect the planet and disk PAs in the same way, such as, for instance, the absolute detector orientation on the sky. On the contrary, all effects that impact differently the planet and disk positions have to be quantified in detail. As the disk and planet absolute PAs may nonetheless be interesting information for other purposes, we address these issues in Sections 3 and 4. We then discuss the relative (projected) position of \bp b relative to the disk. Finally, in section 5, we constrain the de-projected position of the planet, and hence its location within the dusty disk.

\section{Data and reduction procedures}
In our VLT/NACO follow-up program of \bp b, we focused mainly on the determination of the planet orbital parameters. Several images were taken at different epochs. However, the data were either obtained in L' band where the disk is much fainter, and/or with a dithering pattern (different positions of the star on the detector, referred hereafter as different "offsets") to reduce the background (sky and above all detector) noises but limiting the FoV to $\simeq$ 4" (80 AU); such a limited field of view therefore prevented an unambiguous measurement of the main disk component because of the presence of the inner warped component. Without dithering offsets, a much larger field can be obtained, allowing to detect the disk at larger distances, where the main component dominates (the warp component contribution drastically falls down longwards of 80 AU). As a drawback, the background removal is expected to be more critical.
\subsection{The data}
 In order to achieve our goal to measure the  position angle of the planet with respect to the disk using a single set of data, we decided to get VLT/NACO  images at Ks with the star being located at a single offset on the detector.
The observations were carried out on November, 16th, 2010, with the S27 Camera. All data considered here are recorded in Pupil Tracking mode (Angular Differential Imaging, ADI, \cite{marois06}), consisting in a sequence of saturated images (several datacubes) followed and precedented by a series of un-saturated PSF images measured FWHM = 2.98 pixels = 80.49 mas), used to get an estimate of the PSF shape for calibration purposes (photometry, shape), and fake planet simulation (see below). In the present context, we are not so much interested in the photometry but rather in the PSF shape (which is used to inject fake planets, see below). Immediately after the last saturated image, the telescope was offsetted and images with  detector integration times (DITs) similar to those of the saturated data were taken, to estimate the sky/background. 
We tried to observe \bp at parallactic angles such that the telescope spiders did not overlap the disk. 
Finally, the detector plate scale was measured on 5 stars located in an Orion field, using HST astrometric data,  to be 27.01 \pm 0.05 arcsec per pixel, and the North orientation offset  (= absolute PA - PA measured on the detector) was measured to be -0.25 \pm 0.07\deg . These 5 stars are those we use regularly for our astrometric calibrations (\cite{chauvin11}). If, instead of using these five stars, we use 15 stars in the same field, we find the same offset (within error bars): -0.29\deg, but with an increased dispersion (0.3\deg). We attribute the discrepancy in dispersions to possible systematics due for instance to the limited precision of the HST astrometric data which serve as references for the stars positions, and also to the fact that we use in the second approach fainter targets,  for which the  centroid measurements may be less precise. This shows in any case how difficult it is to get precise {\it absolute} PAs and further inforces the relevance of our present approach, which allows to get rid of these uncertainties.
 

\begin{table*}[t!]
  \caption{Ks data used in this paper. 
 ``Par. range'' stands for the parallactic angle at the start and end of 
observations; 
``EC mean'' for the average of the encercled energy and ``t0 mean'' for the average of the coherence time during the observations. }
\label{stats}
\begin{center}
\begin{tabular}{ l l l l l l l l l}
\hline 
 Date & UT-start/end&DIT & NDIT & N exp. & Par. range & Air Mass & EC mean & t0
mean  \\ 
        &(s) &     &     &      \deg &  & &$\%$& (ms)\cr
\hline 
Nov. 15/16, 2010 & 06:02/06:06 &0.11 (ND filter)&100 &12  &-23/-20 &1.13 &41.6&2.3\\ 
Nov. 15/16, 2010 & 06:07/07:25&0.15 & 100&230  & -20.6/13.7&1.13 &35.4&5.1 \\ 
Nov. 15/16, 2010 & 07:29:07:33&0.11 (ND filter)& 100&12  &20.5/22.5 &1.13 &23.5&1.9\\ 

     \hline
    \end{tabular}
  \end{center}
\end{table*} 

\subsection{Data reduction}
As a first step, the data were reduced with different procedures, that we name cADI, cADI-disk, sADI, rADI and LOCI. Indeed, all ADI reduction procedures induce biases that may impact differently the disk or planet final positions; they are then important to know in the present context, to identify the best methods and derive associated error bars.

 The cADI, sADI and LOCI procedures were already described in \cite{lagrange10}. Basically they differ in the way to estimate the star halo (that we will somewhat abusively call "PSF" hereafter) that has to be subtracted from the data to allow planet detection. Briefly, in cADI, the PSF is taken as the mean or median of all individual recentered ADI saturated images\footnote{to recenter the different data, we proceed by fitting (Moffat fitting) the wings of the saturated PSFs  below a threshold of 6500 ADU- see below a discussion on the impact of the threshold value.}; all individual images are then subtracted from this PSF and the residuals thus obtained are derotated and combined (mean or median) to produce the final image. sADI computes "PSFs" for each individual image, using a given number of images taken before or after the considered image, with a field rotation larger than a given angle expressed in FWHM at a given separation (the angle is then constant for all separations). The residuals thus obtained are then derotated and combined to produce the final image. The rADI procedure (identical to \cite{marois06} ADI) is a generalization of the sADI procedure which at each radius, and for each image, once corrected from the median of all images, selects a given number of images that were recorded at parallactic angles separated by a given value in FWHM (the same value in FWHM for each separation), to build a PSF to be subtracted from the image considered. In the LOCI approach, for each given image and at each given location in the image, we compute "parts" of "PSFs", using linear combinations of all available data, with coefficients that allow to minimize the residuals in a given portion of the image.  Finally, cADI-disk is a variation of cADI procedure which after cADI reduction, subtracts to the initial PSF a rotated image of the residuals, in order to remove as much as possible the contribution of the disk to the so-called "PSF". The disk-corrected PSF is then subtracted to the individual images; the individual residuals are then derotated and stacked (median or average) to get the final image. This procedure aimed mainly at checking the impact of the self-subtraction with the cADI procedure. 

In practice, we fixed some parameters for sADI, rADI and LOCI procedures :
\begin{itemize}
\item [-] LOCI: $\Delta r=2\times\,FWHM$ (radial extent of the subtraction zones); $g=1$ (radial to azimuthal width ratio), $N_A=300\times\,FWHM$ (surface of the optimization zone); separation criteria between 1.25, 1.50 and 2.00 $\times\,FWHM$
\item[-] rADI: separation criteria: 1., 1.25, and 1.5  $\times\,FWHM$; number of images used to compute each "PSF" : 10. Note that the criteria that are optimal for the planet detection and those optimal for the disk detection are not the same (see below).
\item[-] sADI: the separation criteria was 1.3, 1.5 and 1.8 $\times\, FWHM$ at a separation of 12, 13, 15 and 19 pixels; number of images used to compute each "PSF" was between 10 and 20. 
\end{itemize}

\subsection{Disk and planet images}
\begin{figure*}
\centering
\includegraphics[angle=0,width=\hsize]{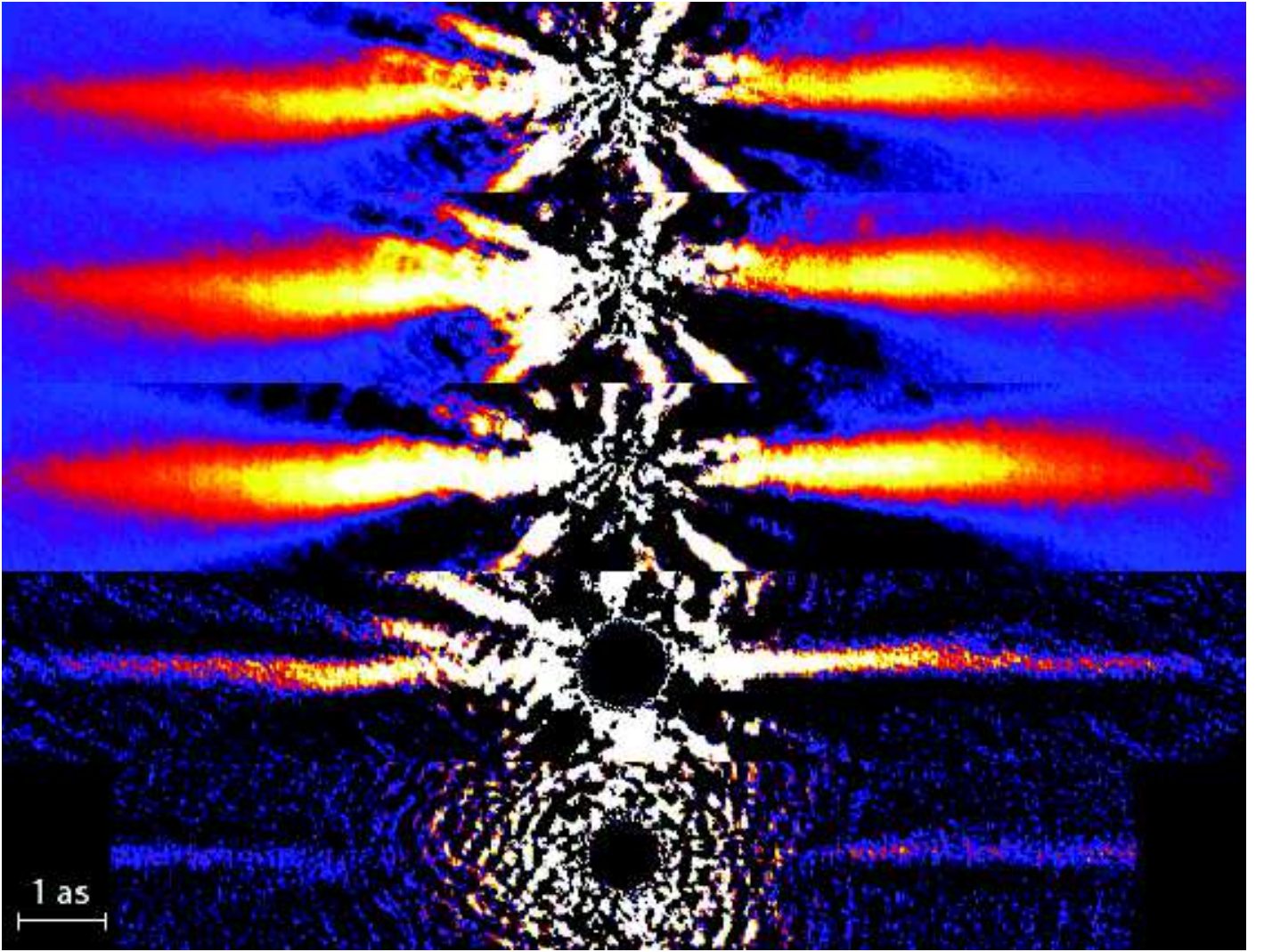}
  \caption{\bp disk at Ks obtained on November, 16th: Ks, S27 data, with a 27 mas/pixel sampling. North-East side is to the left and South-West is to the right. We show the images of the same data reduced with cADI, cADI-disk, sADI, rADI and LOCI. Notes: 1) the color codes are not identical for the different images; 2) the radial thin and bright structures are due to an imperfect removal of the telescope spiders. }
\label{planche_nooffset}
\end{figure*}

\begin{figure*}
\centering
\includegraphics[angle=0,width=\hsize]{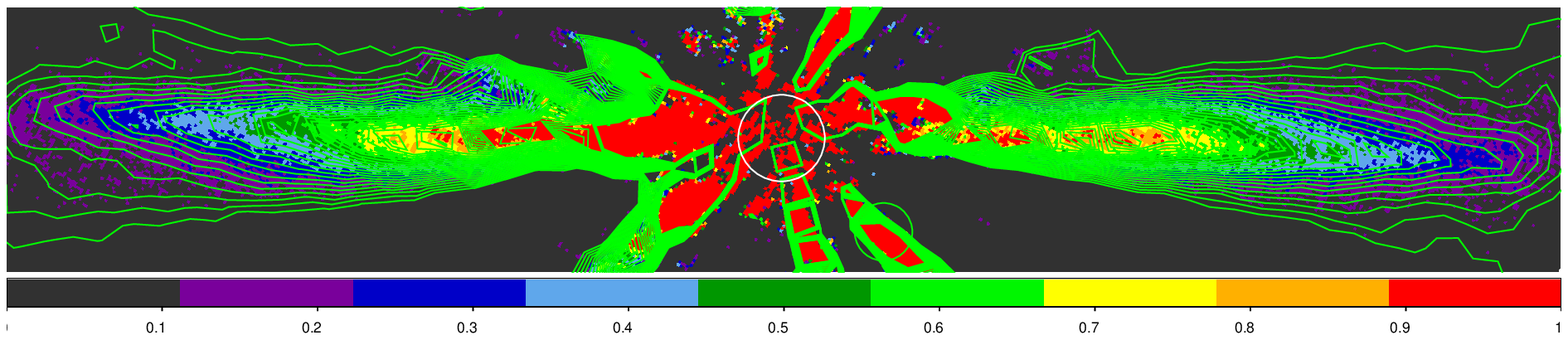}
  \caption{Isophotes (from 0. to 1. ADU) of the  image of the \bp disk at Ks obtained on November, 16th (cADI-disk). North-East side is to the left and South-West is to the right. \bp b is not visible with the present brightness scale, but the white circle, with a radius of 15 pixels indicates the approximate separation of \bp b. }
\label{isophotes}
\end{figure*}

The non-offsetted images reduced with the different methods are shown in Figure~\ref{planche_nooffset}. The disk is well detected out to 130 (SW) and 140 (NE) AU from the star. This allows to identify both the main and warped components.  Figure~\ref{isophotes} shows the isophotes in the case of the cADI-disk image.


The images show a circular pattern located at the NE from the star; this is an artefact that we attribute to the combination of Pupil Tracking observations, PSF saturation and sky removal. This feature strongly affects the detection of the disk at radii shorter than $\sim$2.7" ($\sim$50 AU) at the NE while the disk is clearly seen to the SW at twice shorter radii (see 
Figure~\ref{isophotes}). We will then consider this part as not usable in the following. The artifact is stronger in the non-offsetted images than in the offsetted ones, due to a comparatively less efficient background correction.

If we now compare the disk images produced by the different methods, we see that the disk appears slightly thinner when reduced with sADI, and significantly thinner with rADI than with cADI (or evenmore cADI-disk). The limited impact of sADI on the disk height is due to the fact that the separation criterium (1.5$\times$FWHM) is set at a short separation, and the corresponding angular separation is kept for all  separations, so the disk self-subtraction at large separations ($\geq$ 50 AU) is very limited. This is in contrast with the case of rADI data, where the criterium separation is set at all separations. Disk self-subtraction can in principle be at least partly reduced by choosing larger separation criteria, but in such a case, the image would not be adapted for the planet imaging and position measurement. In the case of LOCI, the disk appears much fainter, and the inclined component is not detectable any longer. This is due to an important self-subtraction, in particular of the inner disk. As a result, we will in the following consider only the cADI and sADI methods to ensure that the disk (and in particular its warped component) is less impacted by the ADI procedures.

\section{Disk orientation(s)}
We first describe different approaches that were adopted to measure the disk(s) orientation(s). An important issue is obviously the uncertainties associated to the measured values. We address this issue in detail in the following subsection.
\subsection{Main disk orientation: measurements}
\subsubsection{Maximum-spine fit to the main disk}
 To estimate the PA of the main disk, we first computed its spine and measured its position angle at separations large enough to ensure that the considered region is dominated by the main disk and that the warp contribution is comparatively smaller. This happens longwards $\simeq$ 80 AU in the present data. HST data (\cite{heap00}; \cite{goli06}) show that indeed, the warped component is about 4 times fainter than the main disk at separations $\geq$ 80 AU. In our data, which have a lower SN at large separations, such a contribution is close to the noise. The considered baseline region for the main disk is then between  $\simeq$ 80 and 120 AU  (160-230 pixels; the outer limit beeing conservatively set by the quality of the data at large radii). 

To define the spine, we first simply identified the brightest pixel in the disk at each radius ("spine-maximum" approach), and measured the PA of the curve obtained in the 160-230 pixels region. To measure this PA, we first took it as a free parameter, and  we derotated the disk around a roughly estimated PA value ($29^{\circ}$  and $209^{\circ}$ resp. for the NE and SW sides, see below) with a step of $0.01^{\circ}$. For each given derotation, we computed the slope of the spine curve; the adopted PA is the angle of rotation for which the slope is null (within \pm 0.01$^{\circ}$). The PA obtained for the NE and SW sides of the disk are given in Table~\ref{maindisk_pa} for the cADI data~\footnote{Note that the values given here are the averages of the values obtained using 3 different data reduction pipelines, in order to minimize the effect of the reduction procedures. In practice, the values obtained were very close to each other, very well within the error bars.}, together with the associated uncertainties which are described in Appendix A. With the rough maximum-spine method, we find with cADI a PA of 29.33$^{\circ}$$^{+0.22}_{-0.30}$ and 209.10$^{\circ}$ $^{+0.22}_{-0.38}$  for the NE side and SW sides respectively, assuming an uncertainty of 0.07~$^{\circ}$ for the True North position. We find very similar results with the median data. We also note that the values found are close to those derived by eye from the isophotes extrema in the same region of the disk: 29.2$^{\circ}$ and 208.9$^{\circ}$ when considering the isophotes between 160 and 230 pixels. 
Table~\ref{maindisk_pa} also provides similar measurements made on the sADI and rADI data, together with associated uncertainties. The sADI data give quite similar values, which is consistent with the fact that given the parameters adopted, the sADI process is not very different from the cADI process at large separations. 
The rADI data lead to lower values (yet coherent with the other given the error bars). We remind that the disk is much more impacted by the rADI process than by the cADI or the sADI ones (with the parameters adopted). 

\begin{table*}[t]
\caption{Main disk position angle (NE/SW), as measured with the various methods described in the text,for cADI, sADI, and rADI (based on averages of resp. 3, 3 and 1 reduced images). In each case, we give the measured values by various fitting methods, and uncertainties associated to these measurements (first line) values, with all biases and uncertainties included (Note that we have assumed here an uncertainty associated to the True North position of 0.07 deg.).}
\begin{center}
\begin{tabular}{lllll}
\hline
&spine, maximum (NE/SW) &               spine, weighted lorentzian (NE/SW)      & 2 cpts fit    (NE/SW) \cr
\hline 
cADI mean&&&\cr
 PA (Main) [$^{\circ}$]& 29.33$^{+0.22}_{-0.30}$ /209.10 $^{+0.22}_{-0.38}$     &29.29$^{+0.13}_{-0.30}$/209.35 $^{+0.14}_{-0.37}$      &29.07$^{+0.20}_{-0.19}$/209.00$^{+0.16}_{-0.15}$ \cr
 warp incl.[$^{\circ}$]&                        NA      &NA                     &3.9 / 3.9 $^{+0.6}_{-0.1}$     \cr
\hline
cADI median&&&\cr
 PA (Main) [$^{\circ}$]& 29.34$^{+0.26}_{-0.28}$/209.11$^{+0.26}_{-0.37}$       &29.28$^{+0.12}_{-0.29}$/209.35$^{+0.12}_{-0.36}$                               &               29.08$^{+0.18}_{-0.18}$/ 209.01$^{+0.14}_{-0.14}$\cr
 warp incl. [$^{\circ}$]&                       NA      &NA                     & 4.1 /3.9 $^{+0.6}_{-0.1}$ \cr
\hline
sADI mean&&&\cr
 PA (Main) [$^{\circ}$] & 29.38  $^{+0.42}_{-0.50}$/ 209.22$^{+0.44}_{-0.44}$ &29.27$^{+0.19}_{-0.35}$  / 209.41$^{+0.25}_{-0.47}$& 29.02$^{+0.30}_{-0.28}$/209.03$^{+0.26}_{-0.24}$  \cr
 warp incl. [$^{\circ}$]&                       NA      &NA                             &  3.5  /3.95 $^{+0.6}_{-0.1}$
\cr
\hline
sADI median&&&\cr
 PA (Main) [$^{\circ}$]& 29.36  $^{+0.50}_{-0.42}$      /209.20 $^{+0.50}_{-0.50}$      &29.23 $^{+0.17}_{-0.32}$/209.39        $^{+0.17}_{-0.39}$                                      &28.98$^{+0.26}_{-0.24}$ /209.02$^{+0.22}_{-0.20}$      \cr     

warp incl. [$^{\circ}$]&                     NA      &NA                             & 3.7/ 3.9 $^{+0.6}_{-0.1}$ \cr
\hline
rADI mean&&&\cr
PA (Main) [$^{\circ}$]& 29.0$^{+0.50}_{-0.36}$ /209.01$^{+0.57}_{-0.43}$ &              29.0$^{+0.37}_{-0.35}$ /209.04$^{+0.44}_{-0.42}$  &  NA \cr
warp incl. [$^{\circ}$]&       NA                      &       NA                                      &NA     \cr
rADI median&&&\cr
PA (Main) [$^{\circ}$]&        29.14$^{0.54}_{-0.40}$/209.04$^{+0.61}_{-0.47}$                    &    29.26$^{0.37}_{-0.34}$/209.07$^{+0.43}_{-0.41}$                       &NA             \cr
warp incl. [$^{\circ}$]&       NA                      &       NA                                      &NA     \cr
\hline
\end{tabular}
\end{center}
\label{maindisk_pa}
\end{table*}

\subsubsection{lorentzian fit to the main disk}
In a second approach, expected to be more precise, we fitted the vertical profile of the derotated disk at each radius, with a weighted lorentzian profile, with a weight proportional to the 4th power of the flux, to enhance the maximum of the profile. We then proceeded as done for the spine-maximum approach. Again the results are given in Table~\ref{maindisk_pa} for the cADI, sADI and rADI data, together with associated uncertainties (see Appendix A). They appear to be quite consistent with the ones obtained with the spine-maximum approach, with yet slightly improved error bars.

\subsubsection{Hybrid fit to the main disk} 
\begin{figure}
\centering
\includegraphics[angle=-0,width=\hsize]{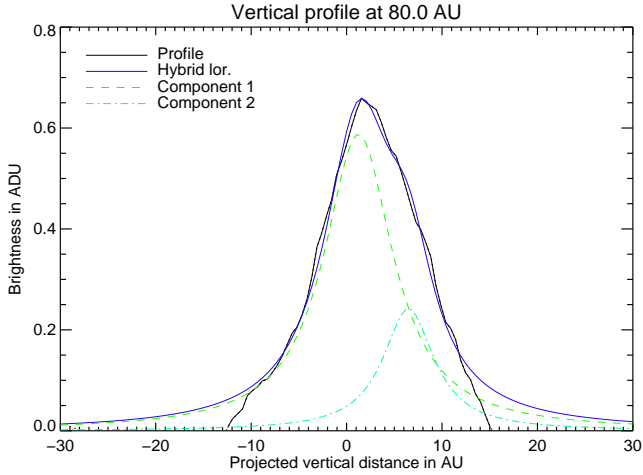}
  \caption{Fit to the vertical profile of the disk (see text) at r=80 AU (NE), by two lorentzians. }
\label{fit_hybride.eps}
\end{figure}

Previous works have fitted the observed disk by a two-component lorentzian profile. This implicitely assumes that the observed disk can be decomposed in two distinct disks \footnote{Note that this may not be the case if the warp is indeed produced by a planet as the warped component shape would not be disk-like (see e.g. \cite{jca01}).}. 
We therefore assumed that the vertical profile of the disk can be fitted by two lorentzian profiles as done in previous works (see e.g. \cite{goli06}), corresponding to the main disk on the one side, and to another inclined disk on the other side. We measured the orientations of both components as follows: we made different rotations of the disk with values close to the main disk PA, and with steps of 0.01$^{\circ}$. For each rotation angle, we fitted the vertical profile by two lorentzians; we determined the rotation angle that nulls the slope of the vertical distance to the midplane of the main disk in the considered reference region, [160-230] pixels (see above). We find for the NE (resp. SW) side a PA of $\simeq$ 29.1$^{\circ}$  and 209.0$^{\circ}$ for the NE and SW sides of the main disk on the cADI data (see Table~\ref{maindisk_pa}). These values are close to the ones found previously, which confirms that the warp component only marginally effects the disk orientation at large separations.  
Once the PA of the main disk was known, we determined the slope of the second component, which gives the tilt of the inclined disk relative to the main disk. An example of a decomposition of the vertical profile by 2 lorentzians at r=80 AU is given in Figure~\ref{fit_hybride.eps}. The values are found to be very similar, but slightly lower ($\simeq$0.3$^{\circ}$) than those derived with the weighted lorentzian and with the spine-maximum methods. We attribute this discrepancy to a small contribution of the warp component in the 160-230 pixels region that impact the results obtained with these last two methods. This is in qualitative agreement with the difference found when considering the isophotes either between 160 and 230 pixels, or further away (between 200 and 240 pixels): 0.4$^{\circ}$ and 0.3$^{\circ}$  for the NE and SW sides of the disk. This is also in agreement with the values found with the weighted lorentzian fit to the disk between 200 and 240 pixel: in such a case,the PA is smaller by 0.2$^{\circ}$ with respect to the PA measured between 160 and 230 pixels. Finally, sADI data provide similar results, and rADI data do not allow such a fitting, due to the important self-subtraction of the warped component.

\begin{figure}
\centering
\includegraphics[angle=0,width=\hsize]{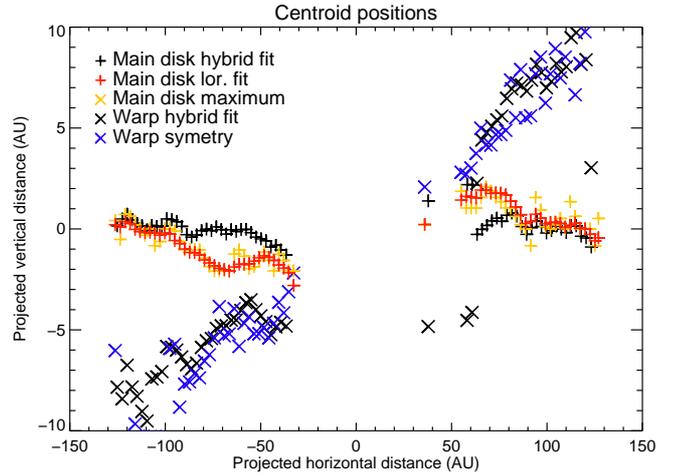}
  \caption{Warp position as estimated with the 3 different approaches (cADI data). NE is to the left and SW to the right. For clarity purpose, we show here the average of 5 consecutive data points along the disk.}
\label{main_warp_position}
\end{figure}

As a summary, Figure~\ref{main_warp_position} shows the results on the warp position with respect to the main disk,  obtained with the different methods. The overall agreement between the three methods is clear; the impact of the warp contribution on the main disk lorentzian or spine fit is also visible shortwards 80 AU. 

\subsubsection{Simultaneous fit to the NE and SW sides}
With all the approaches above, it appears that the NE and SW sides of the main disk are aligned within the error bars. We then made a last estimate of the main disk PA by performing a linear regression considering {\it both sides} of the disk at the same time, hence without assuming a priori a given center for the star. On the cADI data, averaging the lorentzian and 2-component fit results, we find a  PA of  29.17$^{\circ}$ (mean or median), to be compared to 29.18$^{\circ}$ (mean) and 29.17 (median) when considering the NE side of the disk alone, and 209.17 (mean or median) when considering the SW side. sADI measurements are within 0.04$^{\circ}$ of cADI ones. The star center appears to be offsetted by less than 0.05$^{\circ}$ (\pm 0.1$^{\circ}$) from the disk main plane (all error bars included, true North corrected).  

\subsection{Main disk orientation: uncertainties and systematics}
The different identified sources of uncertainties for the disk position angle and their estimations are described in Appendix A. Briefly, we identified two main classes of uncertainties. The first one is related to the data reduction and calibration, and in particular, uncertainties associated with the imperfect determination of the star center on the heavily saturated images, uncertainties associated to the the way the "PSF" to be subtracted during the ADI reduction is estimated, uncertainties related to the disk self subtraction, and uncertainty related to the True North (hereafter TN) position (important only for absolute measurements). The second class is related to the PA measurements themselves: fitting of the disk, and impact of the region considered for the PA determination. Finally, we end up with rather small error bars, of the order of \pm 0.2-0.5 degree (see Table 2). We remind nevertheless that an absolute calibration of the TN would increase the error bars by an additional 0.3$^{\circ}$.

\subsection{Warped component orientation: measurements and associated uncertainties}

We first fitted the vertical profile of the disk by a 2 component profile as described above. However, to optimize the measurement of the warp PA, we chose regions where the warp contribution is more important than previously: [130-160] and [130-180] pixels, i.e.  $\simeq$ [68-83] and [68-94] AU. We then averaged the results (PA values averaged and uncertainties quadratically added) of the different reduced images for each method.  The obtained values are $\simeq$ 3.5-4$^{\circ}$.

 The measurement of the tilt between the warp and main components is free from most of the errors associated to data reduction and calibrations and to the ADI process (assuming, realistically, that they impact similarly the main and warp component). The main sources of uncertainty in this case are rather related to 1) the determination of the warp itself with the 2-lorentzian profile fitting or the symmetry approach, 2) the impact of disk self-subtraction, and 3) the assumptions taken to compute the final image (mean/median). The last source of uncertainty has been found to be negligible. We performed simulations of a 2-component fitting using two fake disks and found that 1) the 2-component fitting approach under-estimates the inclination of the warp component by 
0.5$^{\circ}$ and 2) that the ADI treatment also underestimates the inclination by another 0.1$^{\circ}$. So the measured inclination is probably underestimated by 0.6$^{\circ}$. These values have been taken into account in the error bars provided in Table~\ref{maindisk_pa}.  Note that we chose to consider them as errors rather than offsets, as the present estimation probably depends on the assumptions on the disks geometries.

\begin{figure}
\centering
\includegraphics[angle=-90,width=\hsize]{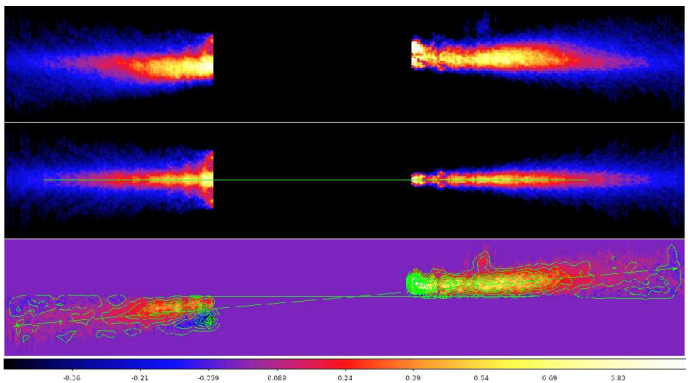}
  \caption{An estimate of the warp. Top: initial image. Middle: estimated main disk. Bottom: subtraction of the first two images, to reconstruct the  warp.}
\label{warp_reconstruit}
\end{figure}

In a second, exploratory approach (referred to as "warp symmetry"), we tried to separate the warp contribution from the main disk one. To do so, for the NE (resp SW) side of the disk, we isolated the contribution of the disk above (resp. below) the midplane and symetrize it to build an estimate of the main disk. We then subtracted this estimate of the main disk to the observed disk to get an estimate of the warp contribution. An illustration of the method is given in Figure~\ref{warp_reconstruit}, and the results obtained are given in Figure~\ref{main_warp_position}. The inclination of the warp thus reconstructed was measured in the  [130-160] and [130-180] pixel regions, with the maximum-spine method or with a lorentzian fit. The obtained values range between 3.6$^{\circ}$ and 4.6 $^{\circ}$ (cADI, sADI), in agreement with the values obtained with the 2-component fit. However, as we do not have a proper way to estimate the biases introduced by this method, we believe that this method should be regarded only as illustrative.
 

\subsection{Conclusions on the disk(s) orientation(s)}
\subsubsection{Adopted values for the disk orientations}
Taking into account the results given in Table 2, we conservatively deduce a PA of 29.3$^{+0.2}_{-0.3}$ degrees for the NE side of the main disk  and 209.2$^{+0.2}_{-0.3}$ degrees for its SW side. Note that we assume here an uncertainty of 0.07$^{\circ}$ for the True North position. We furthermore adopt an inclination of 3.5-4.6$^{\circ}$\  between the warp disk/component and the main disk.

\subsubsection{Comparison with previous results}
There are actually very few published data on the position angle of the disk as observed in the optical or near-IR\footnote{Early sub-mm (850 \mic) data found a disk PA of 32\pm 4$^{\circ}$  at outer distances (\cite{holland98}), while recent 1.3 mm observations indicate a PA of the order of 34$^{\circ}$, but without error bar (\cite{wilner11}).}. In particular, to our knowledge, no value of the main disk PA has been published from HST ACS or STIS data. Kalas and Jewitt (1995) report a PA of 30.1$^{\circ}$ and 211.4$^{\circ}$  for respectively the NE and SW sides. The values vary within a 1 to 2.5$^{\circ}$   range. Noticeably, the data used to mesure the PA were tracing only the outer part of the disk, so they correspond to the main disk PA. More recently, we proposed a revised lower value of 29.5$^{\circ}$ for the position of the main component (\cite{bocca09}). Our measurements are then compatible with the values found by \cite{bocca09} and slightly marginally compatible with those found by \cite{kalas95}. The present error bars associated are much smaller. Finally, we note that HST data (\cite{goli06}) indicate that the NE and SW sides are not perfectly aligned, with a difference of about 0.9$^{\circ}$ (see also their Fig. 10). The present data do not seem to support this result, but we note that given the uncertainty associated to the difference in PA of the NE and SW sides, $\simeq$  0.5$^{\circ}$, one may consider that this conclusion is not very significant. Also, part of the discrepancy might come from the fact that the reference regions used by \cite{goli06} and by us to make the hybrid fit are different: respectively 80-250 AU and $\simeq$ 80-120 AU (see above).

The tilt between the warp component and the main disk has been mesured by HST, with values of $\simeq$ 3$^{\circ}$ (\cite{mouillet97}) using ESO/T3.6m AO ground based data, and 4-5$^{\circ}$ 
(\cite{heap00}), $\simeq$ 5$^{\circ}$ (\cite{goli06}) both using HST, higher SN data. The tilts measured here ($\simeq$ 4$^{\circ}$) are then compatible with previous estimates. However, the comparison is probably not entirely meaningful as we, conversely to \cite{goli06}, constrain the center of both disks to be identical; this impacts then the measured tilt. For instance, it seems, from their Fig. 9, that constraining both centers to be identical would lower the PA of the NE side, without changing significantly the PA of the SW side. The latter further find that the SW part is less tilted (4.7\pm 
0.3$^{\circ}$) than the NE side (5.9\pm0.6$^{\circ}$). The present analysis does not allow to confirm or disprove these results. {Finally, \cite{weinberger03} give a PA of 33.3\pm 2 degrees from mid-IR data. Also, \cite{wahhaj03} appear to adopt the \cite{kalas95} for the main disk PA, but show the 82 AU radius clump as offset from the main midplane by about 2 degrees. It has to be kept in mind that these mid-IR data correspond to thermal emission from the dust where as the near-IR data correspond to scattered light. }

\section{Planet position}
\subsection{\bp b position: measurements}
To measure the planet position, we used the same images (obtained with the same reduction procedures and parameters, and using the positive parts of the residuals) as those used to compute the disk PA. We determined the position of the planet with a centroid calculation using either a 2D gaussian fitting or Moffat fitting of the observed signal. The results are provided in Table~\ref{planet_pos}, together with the associated uncertainties resulting from 
the analysis of all identified sources of errors (see Appendix B). Note that as in the case of the disk, the values given here are the average of measurements performed on images obtained with different reduction softwares (3 different measurements for the cADI data, 3 for the sADI data, and 1 for for the rADI data). 
The planet position is found to be (sep, PA) $\simeq$  (14.2pix; 212.3$^{\circ}$) on the cADI mean. The (sep; PA) values obtained with the cADI and sADI methods differ, but they remain nevertheless compatible given the error bars.  The rADI measurements vary significantly depending on the separation criterium chosen (1.0, 1.2, 1.5, 1.75 or 2$\times$FWHM); this is because when the separation criterium is small, the planet self-subtraction is very important and the residual signal becomes comparable to the noise on the one hand, and when the separation criterium is high, several frames lack comparison frames to build the PSFs and hence the number of frames effectively used to build the final image becomes prohibitively small. They give (sep; PA) slightly lower than the cADI and sADI values, but  still compatible given the error bars.

We note that the various disk PA measurements  showed smaller dispersion than the measurements of the planet position ; this is due to the fact that the measured planet position is very sentitive to the residuals in the final images, where as the disk, located much further away, is not affected by these residuals).


\begin{table}
\caption{Position of bPic b on the different images. Note that we have assumed here an uncertainty associated to the True North position of 0.07 deg. }
\begin{tabular}{lll}
\hline 
cADI mean&(sep, PA) [pix,$^{\circ}$] & (14.23$^{+0.58}_{-0.42}$;212.33$^{+1.12}_{-1.24}$)\cr
cADI median&(sep, PA) [pix,$^{\circ}$] & (14.71$^{+0.26}_{-0.29}$;211.45$^{+1.06}_{-1.21}$)\cr
\hline
sADI mean&(sep, PA) [pix,$^{\circ}$] &  (14.36$^{+0.32}_{-0.65}$;212.27$^{+1.3}_{-1.32}$)  \cr
sADI median&(sep, PA) [pix,$^{\circ}$] & (14.4$^{+0.16}_{-0.43}$;212.35$^{+1.22}_{-1.25}$)\cr
\hline
rADI mean&(sep, PA) [pix,$^{\circ}$] &  (15.05$^{+0.27}_{-0.67}$;212.04$^{+1.27}_{-1.27}$) \cr
rADI median&(sep, PA) [pix,$^{\circ}$]   &  (15.0$^{+0.04}_{-0.48}$;211.92$^{+1.32}_{-1.24}$) \cr
\hline
\end{tabular}
\label{planet_pos}
\end{table}

 In a second step, we injected negative fake planets with variable fluxes and positions (separations and PA) and processed the data to find the position and flux values that minimize the residuals at \bp b location. This method has been described in \cite{lagrange10} and also used in \cite{bonnefoy11}; it is potentially efficient in the sense that it takes into account intrinsically the planet self-absorption during the process (see Appendix B). 
 It appeared that depending on the choice of the unsaturated PSF (either the one taken prior or the one taken after the record of the unsaturated data) used to generate the fake planet signal, the values obtained differ by up to (0.1pix; 0.7$^{\circ}$) in cADI and up to (0.05 pix; 1.25$^{\circ}$) in sADI. Such high differences preclude then the use of this method on the present set of data due to the  important PSF temporal variations that took place during the recording of the saturated images. 
\subsection{\bp b position: uncertainties}
The uncertainties associated to the measurements are described and estimated in Appendix2. They are dominated by the uncertainty on the star position, and by the impact of the low SN of the planet. They are more important than those associated to the disk PA, due to the facts that a) the planet lies within the halo of speckles and b) is close to the star, so the measurements are very sensitive to the uncertainty on the star position.


\section{Relative position of \bp b wrt the circumstellar disk}
\subsection{Planet projected position with respect to the disk}
\begin{figure*}
\centering
\includegraphics[angle=0,width=.45\hsize]{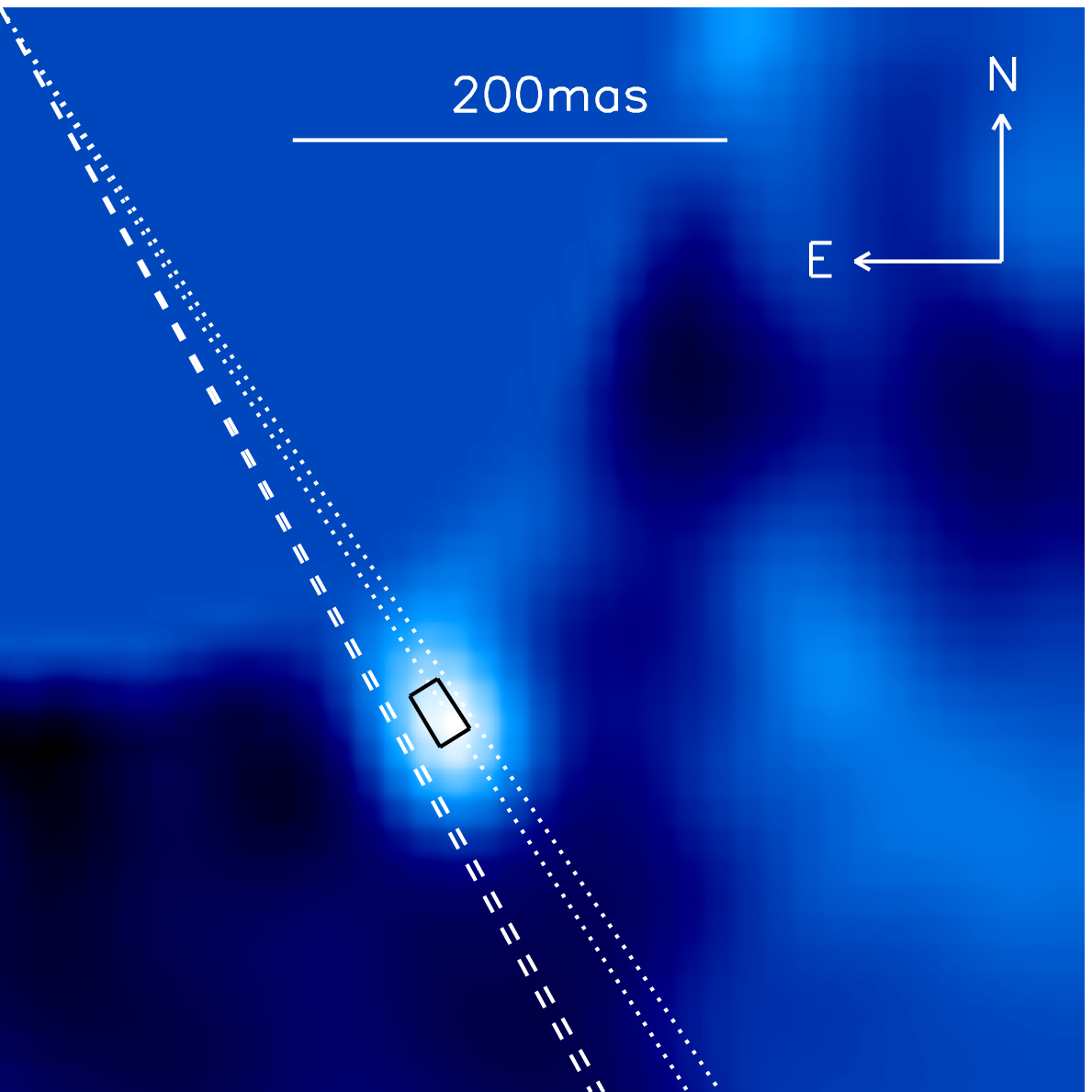}
\includegraphics[angle=0,width=0.45\hsize]{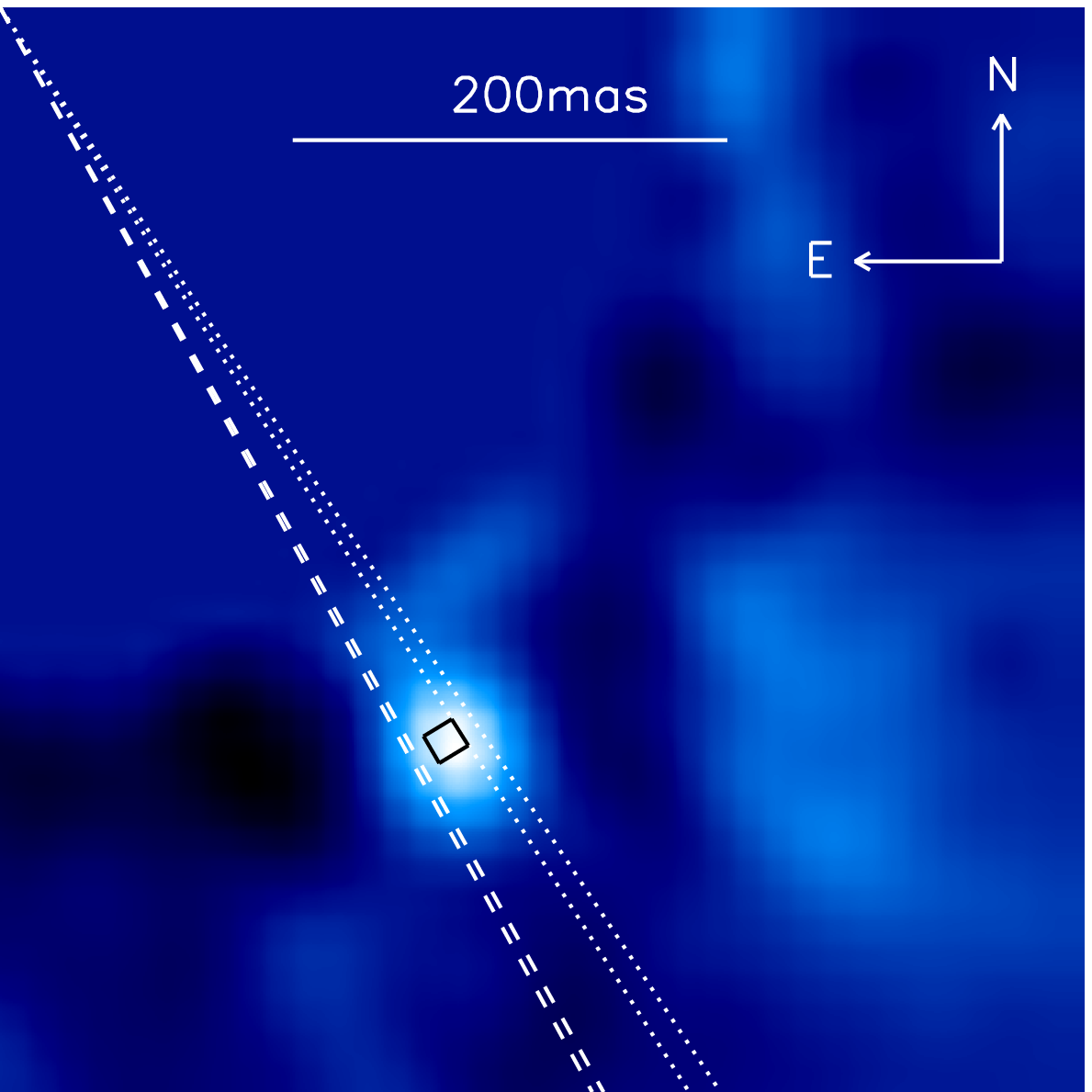}
\includegraphics[angle=0,width=.45\hsize]{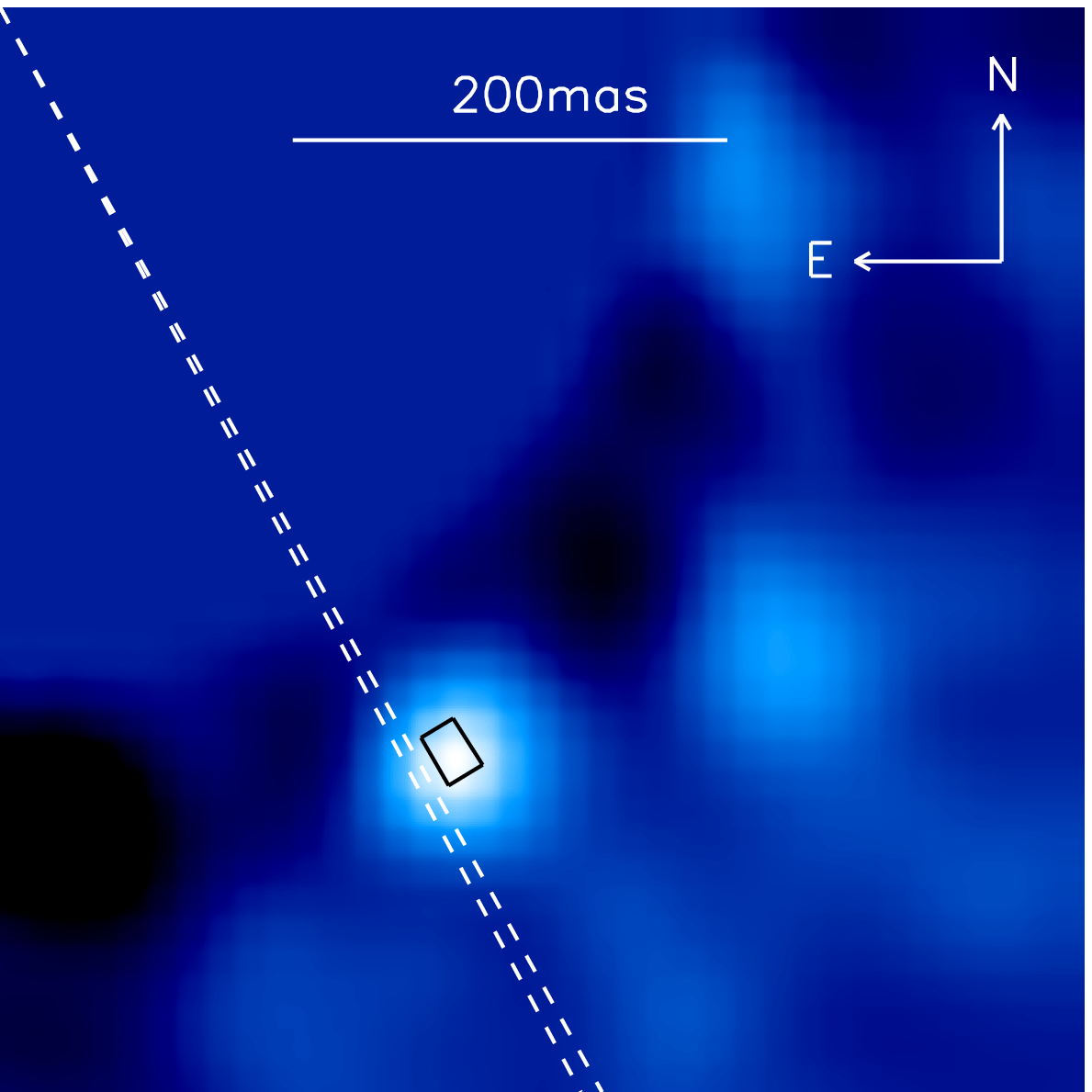}
\includegraphics[angle=0,width=.45\hsize]{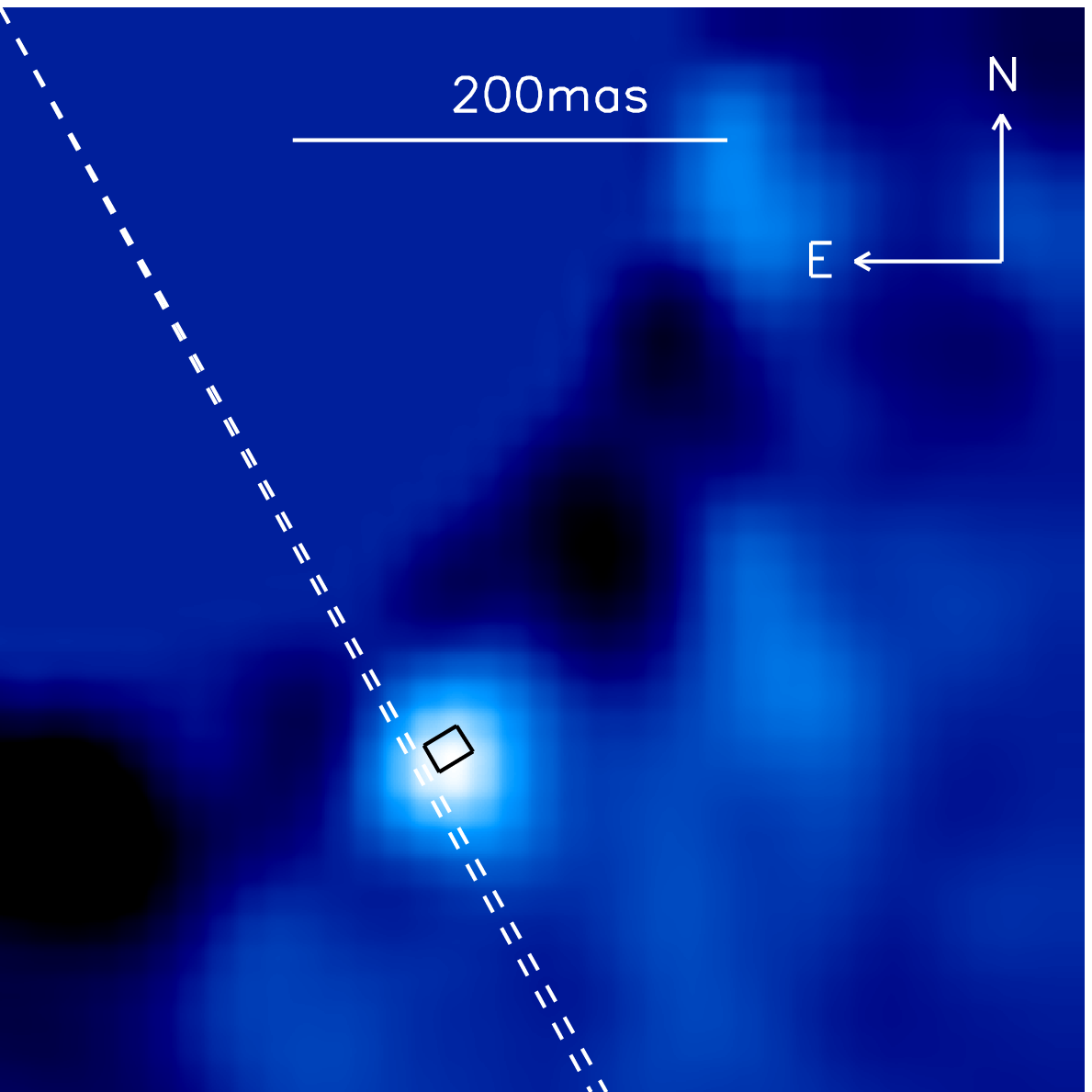}
  \caption{Top: zoom of \bp b position with respect to the disk main component and warp component (mean cADI image), together with associated error bars  pessimistic case}. Left: mean cADI; Right: median cADI. Bottom: idem for rADI images.
\label{image_planet}
\end{figure*}

\begin{figure}
\centering
\includegraphics[angle=0,width=\hsize]{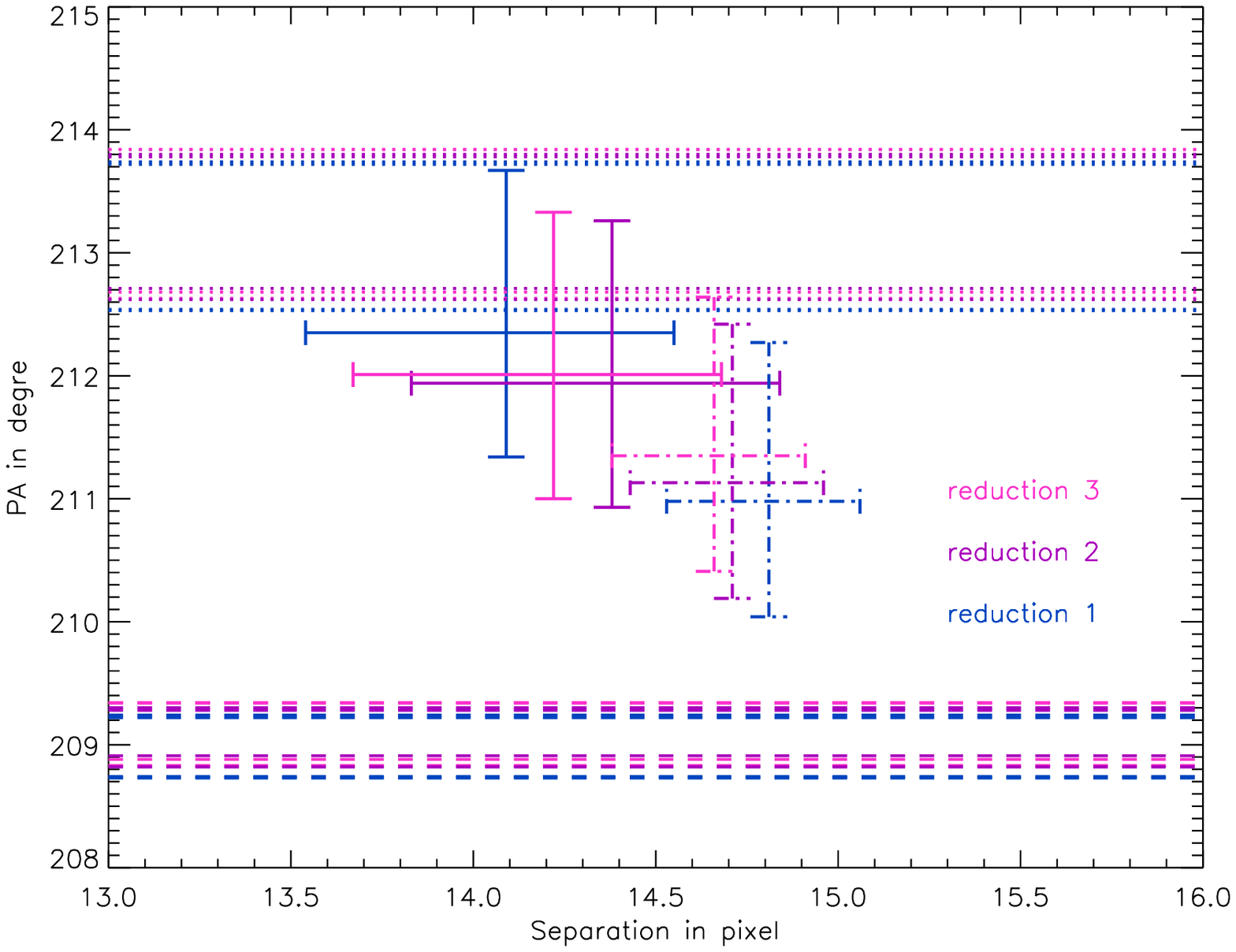}
\includegraphics[angle=0,width=\hsize]{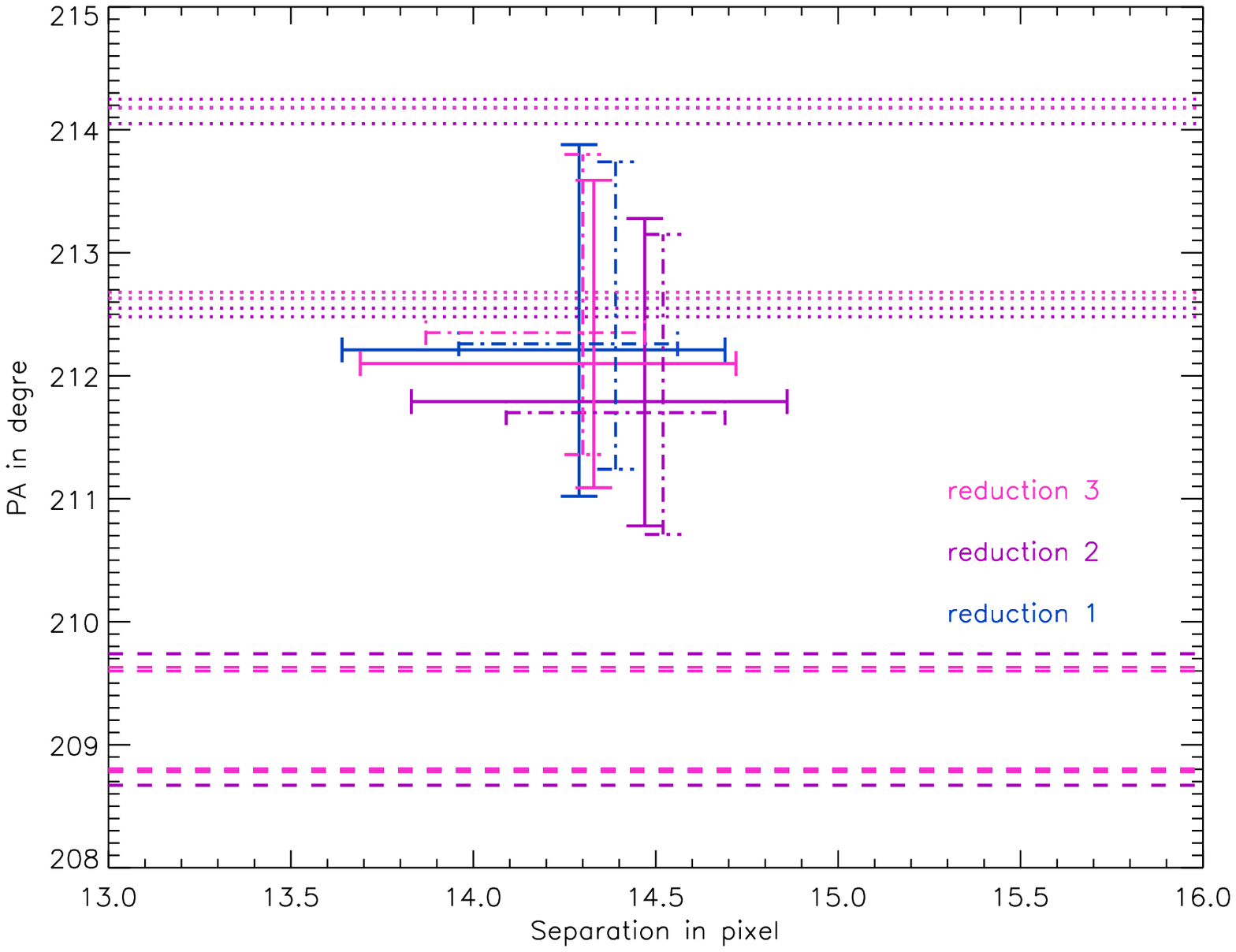}
\includegraphics[angle=0,width=\hsize]{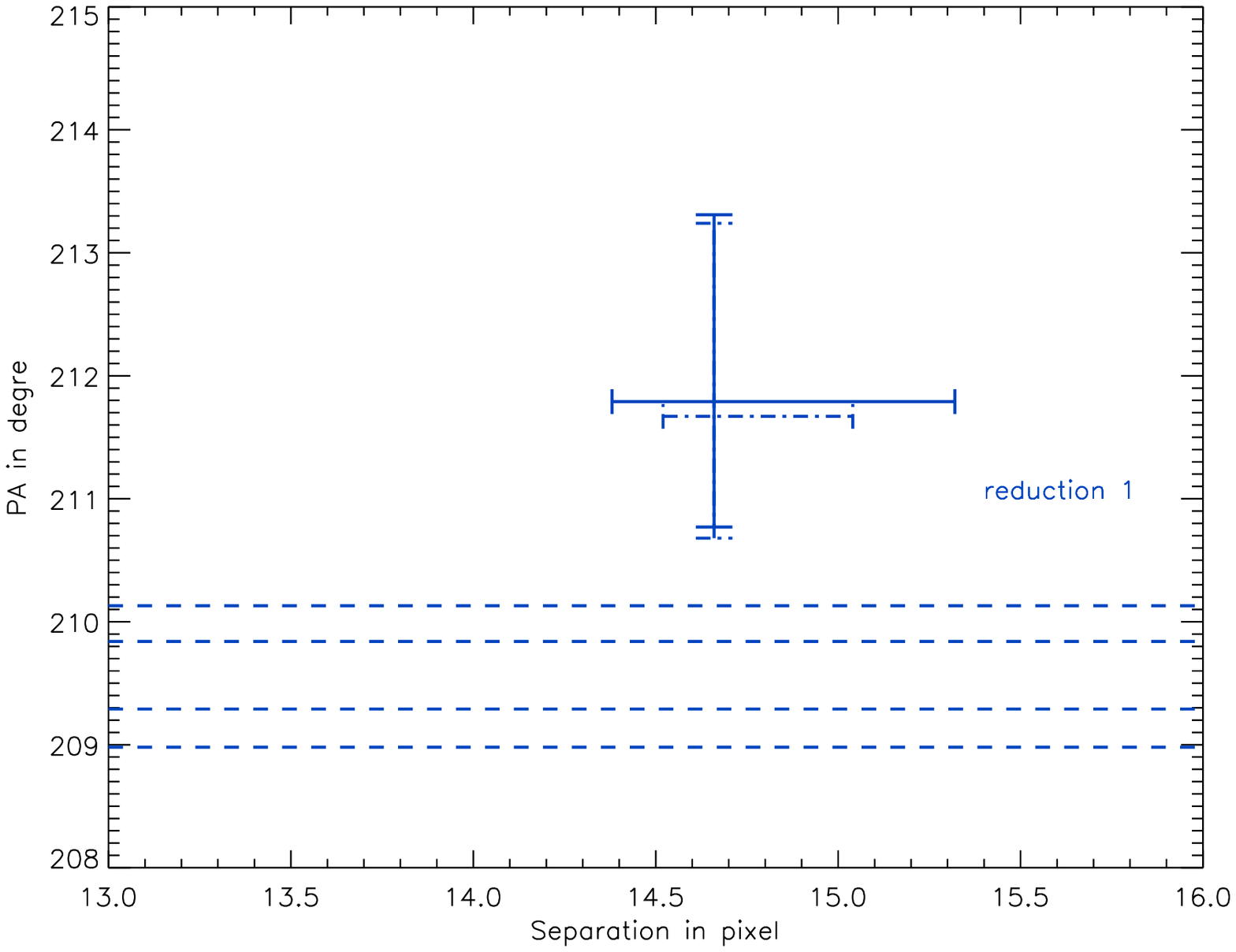}
\caption{Top: position of \bp b together with errors bars measured on 3 different cADI images (mean of the residuals). The full lines correspond to mean cADI data, the dashed lines correspond to measurements on median cADI data. Dashed horizontal lines shows the min and max PAs of the main disk and dotted ones the min and max PAs of the warp component. Conservative error bars have been taken into account. Middle: idem for sADI data. Bottom: idem for rADI data.
}
\label{bilan_position_planet}
\end{figure}

The measurements given in the previous section take into account all systematics/uncertainties. The {\it relative}  position of \bp b with respect to the disk is not affected by the common systematics when a given reduced image is considered (e.g. the uncertainty associated to the True North). Also, the uncertainties depend on the images considered (eg cADI, sADI, mean, median, etc). To explore the planet position relative to the disk, we considered then for each reduced image, the measured disk and planet positions and the associated uncertainties, except the uncertainty associated to the TN. An example is shown in Figure~\ref{image_planet} where we show, for a cADI mean image, a blow up of the planet projected position with respect to the main disk and warped component measured PA. Note that the warped component orientation is the average of the spine-weighted lorentzian fit and the 2-component fit results. \bp b projected position is clearly above the main disk midplane. 
In Figure~\ref{bilan_position_planet}, we summarize all measured individual relative positions on the individual mean (left) and median (right) cADI images, sADI and rADI images as well (note that for rADI, we do not have any value of warp inclination). We see that even though some differences occur from one reduction to the other, all data agree with the fact that \bp b projected position is above the main disk mid-plane.

\subsection{Implication on the position of \bp b wrt the disk}
Our aim is now to use the information on the projected position to constrain the de-projected position of \bp b wrt the disk(s), taking into account the disk(s) orientations.  According to \cite{olofson01}, the gaseous disk rotates towards us in the SW side and from us in the NE side. We assume this is also true for the solid component and the planet. Also we assume,  taking into account the imaging data, that the planet is orbiting with a semi-major axis between 8 and 12 AU from the star (most probable value is 9 AU; \cite{chauvin11}), and has not yet reached the quadrature, i.e. its maximum projected separation from the star. We also assume that the main disk is inclined by 2 to 5 degrees with respect to the line of sight (\cite{kalas95}, \cite{goli06}), and that the nearest-to-Earth part of the disk is tilted above the main disk mid-plane (\cite{goli06}). Under such conditions, if \bp b was located within the main disk midplane, its projected position would be between 0.9 and 2.2$^{\circ}$  (r=8AU), 1.4 and 3.7$^{\circ}$ (r=9AU) and 1.9 and 4.7$^{\circ}$ (r=10AU) {\it below} the observed midplane, which is not consistent with the present observations.

We now assume that the warp component has the same inclination wrt line of sight. In such a case, if the planet is located within the warped component, its projected position would be a few degrees below its midplane, hence between 4 (case of a tilt of 2 degrees between the warp component and the main disk) and 1 degree (case of a 5 degrees tilt) {\it above} the main disk midplane. This is compatible with the present data. 
Given the number of degrees of freedom for the warp characteristics (inclination wrt the disk, wrt line of sight, and inclination wrt line of nodes), we consider that it is not possible to further constrain the planet position.
We note that the MCMC fitting of the astrometric orbit (\cite{chauvin11}) provides ranges of 30-34 degrees for the PA of the planet orbit and 0-3 degrees with respect to the main disk midplane for its inclination. These ranges are in agreement with our present results. 

Overall, these results do not confirm the results of \cite{currie11} who claim that \bp b is located within the main disk. However, as already mentionned, these results were based on the comparison of the measurements of the planet PA with published values of the disk PA (with undefined error bars). The comparison was then affected by different, and probably important systematics.

\section{Summary and future prospects}
Using a single Naco Ks image, we have measured the relative projected position of \bpic b with respect to the disk. The fact that both the disk and the planet projected PA were measured simultaneously removes some possible systematics such as absolute detector orientation, which in turns significantly reduces the uncertainties associated to such difficult measurements. We show that \bpic b projected position is located above the main disk midplane, and actually closer to the warped component. Taking into account our knowledge on the system, we conclude that \bpic b is not orbiting within the main disk, and that the data available today are compatible with an orbital motion within the inclined/warped component. Consequently, \bpic b can be responsible for the inner disk inclination as described in \cite{mouillet97}, \cite{jca01} and more recently in \cite{dawson11}. 
 Future similar observations will be very precious to confirm this result which has important consequences on the disk-planet dynamical interactions. We note that forthcoming high contrast imagers on 8-m class telescopes such as VLT/SPHERE or GEMINI South/GPI, will allow to measure the planet position much more precisely, hence to refine its orbital properties, but given their relatively small FoV, they will not be well adapted for a precise measurement of the main disk PA.

\begin{acknowledgements}
We acknowledge financial support from the French Programme National de Plan\'etologie (PNP, INSU) and from the French National Research Agency (ANR) through the GuEPARD project grant ANR10-BLANC0504-01. This research was also supported in part by the NASA Origins of Solar Systems Program (NNX11AG57G). We thank D Golimowski, G. Schneider and P Delorme for fruitful discussions on the \bp disk and/or on ADI reduction. We finally thank the referee for his/her carefull reading of the paper and comments.
\end{acknowledgements}

\begin{appendix}

\section{Main disk orientation: uncertainties and systematics}
We describe here the different identified sources of uncertainties and how we estimated them. To do so, we either used our actual disk data, or, when needed to get rid of the noise limitations and isolate the impact of a given effect, simulated disk data, with shapes and brightness profiles similar to that of \bpic disk\footnote{We assumed a radial profile density that follows a power law distribution with $\propto$r$^{-4.5}$ further than 102 AU, $\propto$r$^{-2}$ in the 30-102 AU region and $\propto$r$^{2}$ within 30 AU. The vertical structure of the disk is given by: 
$$ I_{vertical}(r,z) = e^{-\left(\frac{|z|}{\xi}\right)^\gamma}$$ 
where the height scale $\xi=\xi_0 \left(\frac{r}{R_m} \right)^\beta$ is 2 AU at 102 AU, and the disk flare coefficient is $\beta$=1.5. The disk brightness is normalized at K=11.5 at 100 AU, corresponding to a disk 10 times brighter than the actual \bp disk, to avoid any error due to the limited signal-to-noise of the real data, and identify properly the bias due to the procedure only. Finally the disk is slightly inclined (i=87.7$^{\circ}$, but this inclination has no consequence on the results). The simulated disk is added to the actual data, at a PA similar to the measured main disk PA., but significantly brighter. } We quantified the impact of the various effects on the cADI, sADI and rADI data (mean/median), whenever the uncertainties depended on the way the images were obtained.


\subsection{Uncertainties and systematics related to data reduction and calibration } 
\begin{itemize}
\item[-] Uncertainty associated with the imperfect knowledge of the star center on the saturated images: this is due to the fact that the NaCo PSFs are not perfectly axisymmetrical and the position of the star center in the saturated images cannot therefore be straightforwardly retrieved from the center of the PSF wings. Ideally (perfectly stable PSF), the offset between the true center and the one estimated from the PSF wings should be treated as an offset (bias). However, due to observed variations between the unsaturated PSFs images taken prior and after the saturated images, we will conservatively rather take it as an uncertainty. This uncertainty impacts both the recentering offsets that have to be applied to each saturated frame,  and hence the center of derotation of PT images (which is also the center of reference for the PA measurement) and finally the measurement of the PA. To estimate the impact on the measured PA, we used a bright fake disk. We first estimated the error associated to the star center position. To do so, we computed the star center (Moffat fitting) on the unsaturated PSFs recorded just prior and after the set of saturated images, once scaled (through the DITs and neutral density filter) to the same flux levels as the saturated PSFs, using either the whole flux range of the PSF, or using only its wings up to varying levels up to 14000 ADU (which is the level of saturation with the given observing mode). It appears that for levels between 5000 ADU and 8000 ADU (which frame the 6500 ADU threshold used for our saturated PSF fitting), the centers offsets are in the range [-0.026; 0.26] pixel on the x-axis, and [-0.011; 0.18] on the y-axis. To estimate the impact of this uncertainty on the star center on the final PA measurements, we used simulated bright fake disks that we added to the actual datacubes (hence in PT mode), and processed these data assuming that the actual star center is shifted by values between -0.26 and +0.026 pixel on the detector x axis and between -0.18 and 0.011 pixel on the y axis; we then measure the resulting disk PA as described above. We find that the impact on the PA is rather small, less than -0.05$^{\circ}$ and 0.03$^{\circ}$. 

\item[-] Uncertainty associated with the way we estimate the "PSF" to be subtracted (either mean or median of the saturated images). This uncertainty was measured on the real data. No detectable impact on the PA measurement was found, which is coherent with the fact that the measurements are performed at large separations.

\item[-] Uncertainty associated with the self-subtraction of the disk in the ADI procedure: this error can be estimated using bright fake disks only; it was then measured on bright disks, and found to be less than 0.01$^{\circ}$ in cADI and sADI, and larger (0.04$^{\circ}$) in rADI (1.2 and 1.5$\times$FWHM), but the latter value strongly depends on the rADI parameters and on the considered region.


\item[-] Uncertainty associated with the determination of the True North (absolute calibration) on the detector. This error was estimated to be 0.07$^{\circ}$ when considering five stars in the field of view. As mentioned before, using all stars leads to a larger dispersion, 0.3$^{\circ}$. In Table~\ref{maindisk_pa}, the error considered is 
0.07$^{\circ}$, which has to be kept in mind.

\end{itemize}

\subsection{Uncertainties and systematics related to the PA measurements} 
\begin{itemize}
\item[-] Uncertainty associated with the fitting of the main disk: for the disk PA, we first estimated the noise level (as a function of star distance) in a disk free, 10$^{\circ}$   angular region. 
For the maximum method, for each vertical profile, we defined the error in pixel of the brightest pixel as the largest vertical distance between pixels having a flux greater than the maximum flux - the noise rms level. For the lorentzian fit, the use of weights prevents from performing a linear fit to the data associated with errors. We then conservatively considered a +/-0.5 pixel error on the center measurement. 
Finally, for the 2-component fit, we took the following approach: for each pixel of a given vertical profile, we associated an error in ADU as the noise rms level at the appropriate distance to the star. In a second step, we derived the error in pixel associated to the position of the center of the main and warped components. This error was then taken into account in the linear regression that was used to compute the slopes (and hence the disk(s) PA) and associated errors. Of course, the estimates were made on the real data. Typically, the error found are less than 
0.08$^{\circ}$.
\item[-] Systematics associated with the region considered for the PA determination: we performed several measurements of the disks PA with reference regions variable in size and positions (within realistic values). On the cADI data (real data), we considered 2 other reference regions, one from 130-230 pixels, and one from 200-240 pixels, in addition to the former one: 160 to 230 pixels. With the 130-230 pixel region, we find a somewhat higher value for the PA than with the reference region, when assuming a single component disk and a similar value when considering a 2-lorentzian fit (Figure~\ref{region_reference}). This is due to the fact that this region is more contaminated by the warped contribution than the [160-230] pixel one. When using the 200-240 pixel region, we find again a similar value with the 2-lorentzian model, and values slightly lower (0.05$^{\circ}$) when considering a single component. The values adopted to build up the error budget are those derived from the comparison of the [160-230] pixel and [200-240] pixel regions only.
 \end{itemize}

\begin{figure}
\centering
\includegraphics[angle=0,width=\hsize]{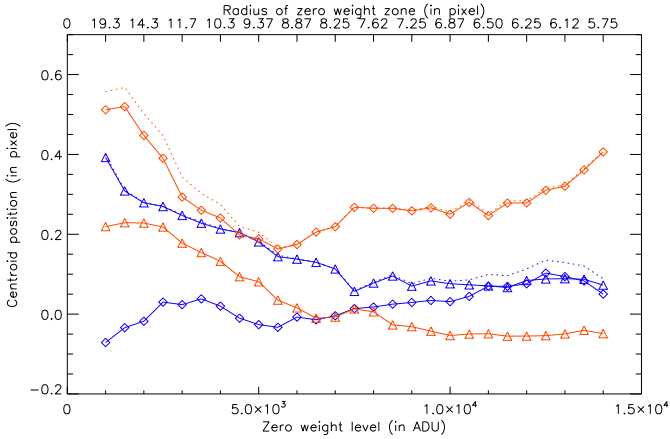}
  \caption{Measured star center on the x-axis (diamonds) and y-axis (triangles) on the detector when considering the whole unsaturated PSFs taken before (red) and after (blue) the saturated images and when considering various radius as thresholds (see text). }
\label{centers_estimates}
\end{figure} 

\begin{figure*}
\centering
\includegraphics[angle=-90,width=\hsize]{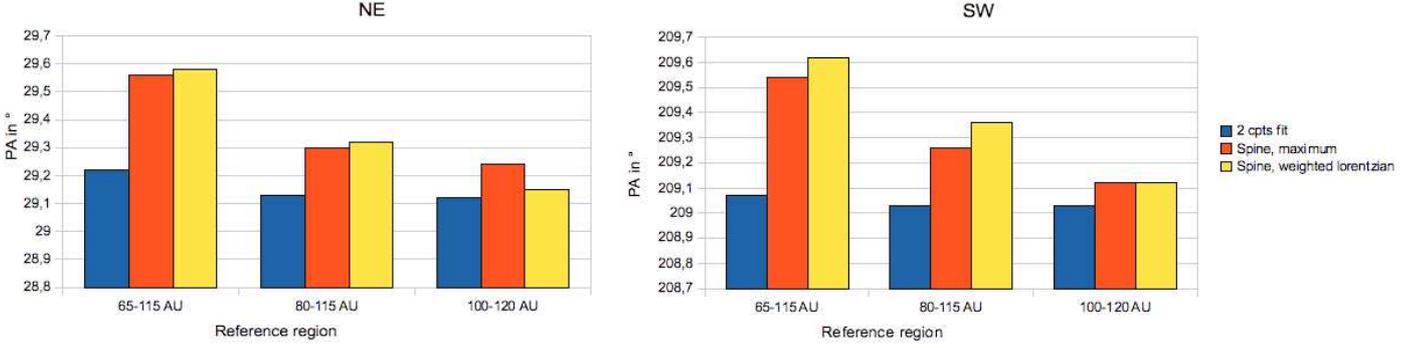}
  \caption{Impact of the considered regions for the estimation of the Main disk PA (cADI images, see text). Three regions are considered: [130; 230] pixels [65-115] AU,  [160; 230] pixels [80-115] AU,  [200; 240] pixels [100-120] AU . Blue:  2-component fit. Red=spine, maximum. Yellow:  weighted lorentzian.}
\label{region_reference}
\end{figure*}

\section{\bp b position: associated uncertainties}
 \subsection{Uncertainties related to the data reduction and calibration} 
To estimate the uncertainties/systematics, we procedeed as for the disk (see above), using bright fake planets positioned at the \bpic b location instead of bright simulated disks, or using the data themselves. The fake planets were built using the unsaturated PSFs\footnote{We checked that taking either the PSF taken prior to the saturated images or the average of the PSFs taken prior and after the saturated images does not change the results, which is due to the fact that our fake planets are bright.} properly scaled in flux.
\begin{itemize}
\item[-] Uncertainty associated with the imperfect knowledge of the star center on the saturated images (see above for a detailed description): we used here bright fake planets injected at the \bp b position to measure this uncertainty, and assumed the same star center offsets as for the disk. The impact on the planet position is found to be quite important, up to 0.3 pixel for the separation, and 0.6$^{\circ}$ for the PA. This contrasts with the low impact on the disk PA and is explained by the fact that the planet is much closer to the star than the disk.
\item[-] Uncertainty associated with the recentering of the individual saturated images with respect to each other within a cube. This occured only in one reduction where the data were intentionnally not recentered within each cube, but were directly collapsed. We find an error of 0.1$^{\circ}$ (resp. 0.3$^{\circ}$) for the measured PA for cADI (resp. sADI).  \par
\item[-] Uncertainty associated with the estimation of the "PSF" to be subtracted (either mean or median of the saturated images): when bright fake planets are considered, the impact is found to be very small (lower than 0.01 pix on the separation and 0.08$^{\circ}$ on the PA) on all data. 
\item[-] Uncertainty associated with the self-subtraction of the planet in the ADI procedure: the impact on a bright fake planet is quite limited on the cADI data ; less than 
0.1$^{\circ}$ for the PA. The effect on fainter fake planets would be more important as the signal is closer to the noise.  We therefore consider this uncertainty below.
\item[-] Uncertainty associated with the residual noise: this uncertainty is the most difficult to measure. Ideally it has to be measured at the location of the planet. We could not find a way to do so because of the presence of the planet. We therefore considered several fake planets, with a flux and a separation identical to those of \bp b, 
 at different PA. We then measured the dispersion of the errors between the position of the injected planets and the actually measured locations after reduction. This leads to quite large uncertainties, up to (0.3pix; 0.5$^{\circ}$) in cADI and (0.3pix; 0.7$^{\circ}$) in sADI. These values are probably conservative as we are considering different directions, and in particular regions which are not free of spiders signatures, which induce higher levels of noise. 


\item[-] Uncertainty associated to the determination of the True North (absolute calibration) on the detector. The conclusions are the same as for the disk. Note that we considered here an 0.07$^{\circ}$ uncertainty.
\end{itemize}

\subsubsection{Uncertainties related to the PA measurement} 
\begin{itemize}
\item[-] Uncertainty associated with the fitting of \bp b: we first checked that using Gaussian or Moffat fitting does not induce significant differences. For Moffat fitting, we made several measurements using apertures with different sizes (from 5 to 7 pixels in diameters), and also we positionned the center of the aperture at variable positions, within 1 pixel of the \bp b estimated center. We find differences up to (0.1pix; 0.2$^{\circ}$).
 \end{itemize}

\end{appendix}

\end{document}